\documentclass[sigconf]{acmart}
\usepackage{graphicx} 
\usepackage{algorithm}
\usepackage{algcompatible}
\usepackage{algpseudocode}
\usepackage{multirow}
\usepackage{booktabs}
\usepackage{makecell}
\usepackage{pifont}
\usepackage{indentfirst}

\newcommand{\cmark}{\ding{51}}
\newcommand{\xmark}{\ding{55}}
\usepackage{xcolor}



\usepackage[dvipsnames]{xcolor}
\usepackage[most]{tcolorbox}
\newtcolorbox{rqbox}[1][]{
    enhanced,
    breakable,
    colback=gray!3,             
    colframe=gray!45,             
    fonttitle=\bfseries,
    boxed title style={         
        colback=gray!65,          
        sharp corners           
    },
    attach boxed title to top left={
        xshift=3mm,             
        yshift=-2.5mm            
    },
    title={#1}                  
}

\newtcolorbox{txtbox}[1][]{
    enhanced,
    breakable,
    colback=gray!3,             
    colframe=gray!45,             
    fonttitle=\bfseries,
    boxed title style={         
        colback=gray!65,
        boxrule=0pt,          
        arc=2pt,
        left=4pt,right=4pt,top=2pt,bottom=2pt           
    },
    attach boxed title to top left={
        xshift=6pt,             
        yshift=-2mm            
    },
    title={#1}                  
}

\renewcommand\footnotetextcopyrightpermission[1]{}      
\tcbset{before skip=4pt, after skip=4pt}

\title{FPGAgent: An LLM-Assisted Framework for Autonomous HLS Code Generation and Verification in FPGA Environments}

\newcommand{\sjtuAffiliation}{%
  \affiliation{%
    \department{Shanghai Key Laboratory of Scalable Computing and Systems}
    \institution{School of Computer Science, Shanghai Jiao Tong University}
    \city{Shanghai}
    \country{China}
  }
}

\author{Tianyu Wang}
\sjtuAffiliation

\author{Wenjie Wang}
\sjtuAffiliation

\author{Jianguo Yao}
\sjtuAffiliation

\author{Haibing Guan}
\sjtuAffiliation

\author{Xijun Li}
\authornote{Corresponding author.}
\sjtuAffiliation
\email{lixijun@sjtu.edu.cn}

\date{February 2026}

\begin{abstract}
\bfseries
Large language models (LLMs) have shown substantial promise for high-level synthesis (HLS) code generation. However, most existing approaches focus primarily on validation at the simulation or synthesis stage. Due to inherent timing and place-and-route constraints in hardware environments, \emph{HLS code that passes simulation and synthesis does not necessarily translate into deployable, runnable designs on real FPGA platforms}. In addition, the lack of publicly available benchmarks has led many methods to be evaluated only on small, self-curated test suites, making their effectiveness difficult to validate. To address these issues, we propose FPGAgent, a multi-agent framework tailored to real FPGA environments, which targets autonomous HLS coding with end-to-end executability validation. To the best of our knowledge, FPGAgent is \emph{the first task-specification-to-executable HLS generation framework that has been experimentally validated on a well-established benchmark}. Given a natural-language task specification, FPGAgent first injects HLS-specific knowledge into the model, and employs evolutionary search to iteratively derive HLS kernel implementations with reliable performance. It then generates a C++ validation program to verify the functional correctness of the produced kernel, analyzes potential functional defects, and guides the model to apply targeted repairs. Finally, it synthesizes host code to enable compilation and board-level execution on FPGA hardware. We comprehensively evaluate FPGAgent using five well-established models on HLS-Eval, a benchmark comprising 78 tasks across multiple domains, and verify board-level executability on real FPGA platform. Compared with existing baselines, FPGAgent improves synthesizable rate by 16.9\% on average, executability by 26.7\%, and functional correctness by 30.6\%. These results indicate that FPGAgent substantially enhances the practical usability of LLM-based HLS generation, demonstrating the practical utility of end-to-end LLM-assisted HLS generation.
\end{abstract}

\begin{document}

\maketitle

\vspace{-0.6em}

\section{Introduction}
Large language models (LLMs) have recently demonstrated strong performance across a wide spectrum of code-centric tasks, including code generation, completion, and repair. In parallel, there is growing interest in leveraging LLMs as a foundation for automating design workflows. In the hardware design domain, this trend has spurred extensive research on LLM-based code generation and verification~\cite{blocklove2023chip,firouzi2025chipmnd,chang2024data}, with the overarching goal of reducing the human effort required to write, optimize, and validate hardware implementations~\cite{liao2024llms,yin2024ado,zhang2025analogxpert,thakur2023autochip,wang2024chatcpu}. Early efforts in this area primarily focused on HDL (Hardware Description Language) code generation and benchmarking, leading to the development of influential datasets, models, and evaluation protocols~\cite{liu2023verilogeval,thakur2024verigen,lu2024rtllm,fang2025nettag,vitolo2024natural,chang2025data}. Subsequent studies have further advanced benchmarking and survey works for larger-scale RTL (Register Transfer Level) designs~\cite{pinckney2025comprehensive,allam2024rtl}.
Although the above studies have helped delineate the initial problem space for applying LLMs to hardware design, recent research increasingly suggests that, compared to directly generating RTL, high-level synthesis (HLS) may serve as a more suitable interface for LLM-driven design automation~\cite{pinckney2025comprehensive,allam2024rtl,liu2024rtlcoder,wang2025verireason,liu2024openllm}.\par

The shift toward HLS is motivated by both representation and tooling considerations. Compared with RTL, HLS code is closer to software-style programming abstractions, making it easier for LLMs to map from natural language task descriptions or C/C++ programs to synthesizable hardware-oriented code~\cite{gai2025exploring,collini2025c2hlsc,xiong2024hlspilot}. At the same time, HLS toolchains (e.g., Vitis HLS) provide structured compiler, simulation, and synthesis feedback that can be directly incorporated into iterative optimization and repair loops~\cite{abi2025hls,moraru2026laafd,mashnoor2025timelyhls}. This combination makes HLS an attractive intermediate layer for LLM-based hardware generation: it lowers the semantic gap between specification and implementation while preserving a practical path to hardware realization~\cite{gai2025exploring,swaroopa2025evaluating,liao2024llms}.\par

\begin{figure*}
    \centering
    \includegraphics[width=0.85\linewidth]{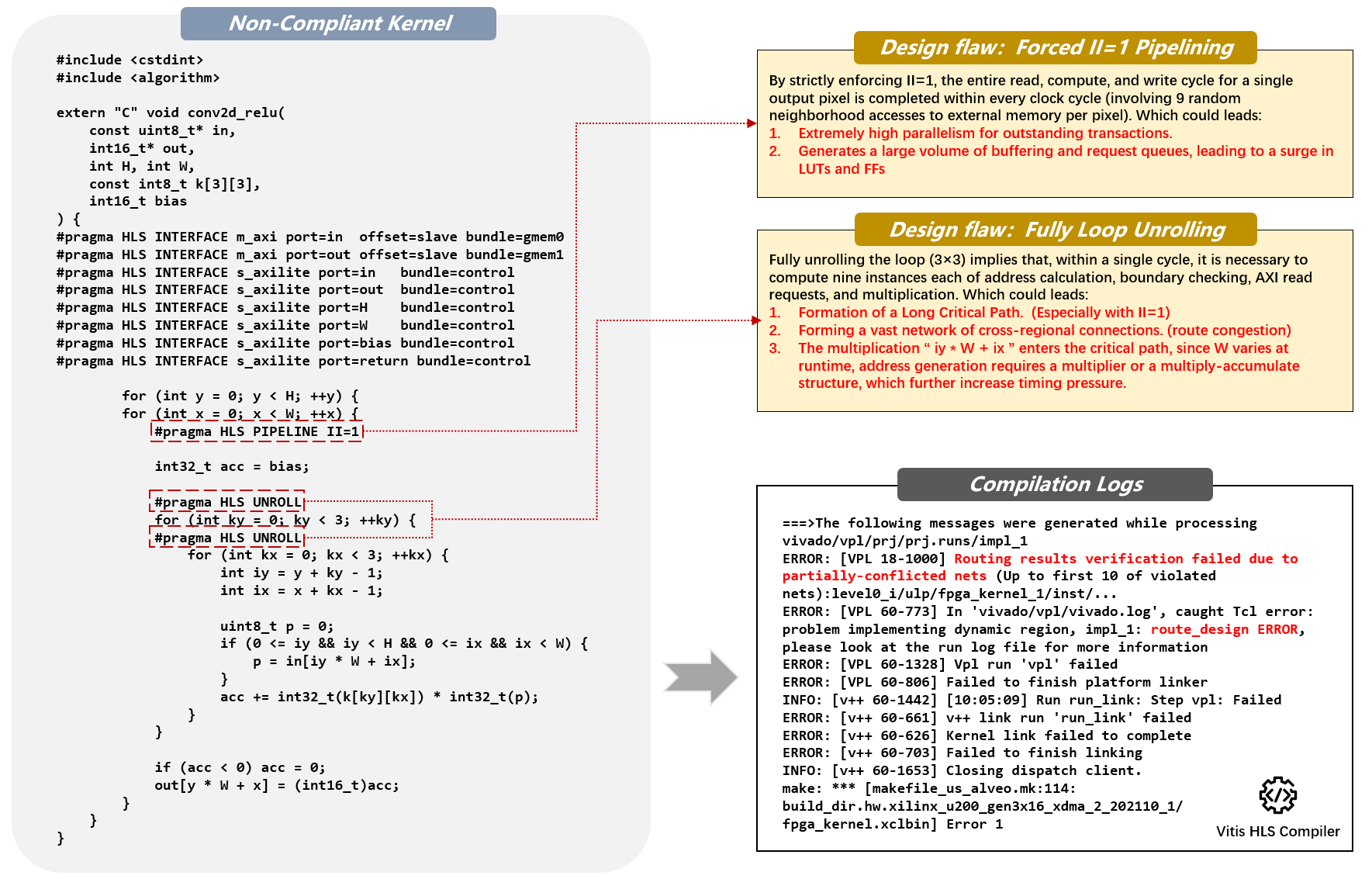}
    \caption{We present a motivating example regarding FPGA kernel code generation: while the code successfully passes simulation tests, its improper pipelining and overly aggressive loop unrolling result in an excessively long critical path within a single clock cycle, consequently leading to routing failures during the generation of the FPGA executable.}
    \label{fig:motivating example}
\end{figure*}

Recent work (especially from 2025 onward) has significantly advanced the LLM-for-HLS landscape along several complementary directions~\cite{xu2025hlstester}. One line of research focuses on HLS code generation, refactoring, and repair, aiming to bridge the software-hardware gap by converting natural language or general-purpose C/C++ code into HLS-compatible implementations. Representative examples include HLSPilot~\cite{xiong2024hlspilot}, C2HLSC~\cite{collini2025c2hlsc}, HLSRewriter~\cite{xu2025hlsrewriter}, and LLM-based HLS program repair methods~\cite{xu2024automated,}, which collectively improve HLS compatibility and reduce the burden of manual rewriting. Another line targets directive/pragma optimization and QoR (Quality of Results) enhancement, where LLMs are integrated with graph-based learning, retrieval augmentation, and design-space exploration to improve timing, resource usage, and performance~\cite{mashnoor2025timelyhls,prakriya2025lift,yao2025high,xu2024optimizing,reddy2025lhs}. In parallel, benchmark and infrastructure efforts such as HLS-Eval and Bench4HLS provide standardized pipelines for evaluating generation quality through parsing, compilation, simulation, synthesis, and analysis~\cite{khan2026bench4hls,abi2025hls,gai2025exploring}. More recently, agentic HLS has emerged as a promising paradigm, where multiple agents or tool-aware iterative loops leverage compiler/synthesis feedback to refine code and optimization decisions over multiple rounds~\cite{moraru2026laafd,collini2025can,zhang2026pragmas}.\par 

Despite these advances, existing approaches primarily concentrate on kernel generation and validation within the HLS toolchain (e.g., compilation, co-simulation, and synthesis)~\cite{khan2026bench4hls,abi2025hls,moraru2026laafd}, \emph{while paying comparatively less attention to whether the resulting designs can be successfully compiled, deployed, and executed on real FPGA platforms}. Given the inherent constraints of hardware environments—particularly those related to timing closure, place-and-route, and resource utilization—passing simulation and synthesis alone does not guarantee that a design is runnable on physical hardware. Moreover, many methods are evaluated only on a small set of self-curated benchmarks, with limited empirical validation on medium- to large-scale benchmarks~\cite{khan2025sage,mashnoor2025timelyhls}. \emph{\textbf{Taken together, existing approaches exhibit multiple limitations: (1) The generated HLS code is not executed and validated on real hardware platforms; (2) Evaluation on well-established benchmark suites is often missing, making it difficult to assess generalization; (3) Some methods focus on optimizing or translating from existing C/C++ implementations, which prevents direct HLS generation from task-level specifications.}} These limitations hinder their applicability to practical HLS development, posing substantial challenges to leveraging LLMs for assisting FPGA engineers. This motivates the need for a more practical end-to-end framework for HLS generation and validation.\par

To address this gap, we propose FPGAgent, a multi-agent end-to-end framework for autonomous HLS code generation and verification targeting real FPGA platforms. Given a high-level task description in natural language, FPGAgent automatically generates HLS kernel code and invokes the industry-standard toolchain to perform simulation, synthesis, and executable construction. In a word, this paper makes the following contributions.\par

(1) We propose a multi-agent, end-to-end HLS code generation and verification framework that maps task design to FPGA executables, and further validate the executability of the generated designs on real hardware. It is the first time to verify an end-to-end HLS generation framework on a well-established benchmark.\par
(2) We develop an iterative, evolutionary HLS code generation module that integrates domain priors with online learning during the design process, and incorporates a multi-level feedback structure to enable parallel exploration of performance optimizations across multiple candidate directions.\par
(3) We propose a module for verifying and repairing the functional correctness of HLS code. This module validates functional behavior at C/C++ level, diagnoses potential functional defects, and drives the model to apply targeted repairs.\par
(4) We conduct comprehensive comparative and ablation studies on HLS-Eval~\cite{abi2025hls}. We establish multiple evaluation metrics to quantify the effectiveness of the FPGAgent framework and provide an in-depth analysis of the contributions of its individual components.\par

The structure of this paper is as follows. Section 2 presents our motivating example. In Section 3, we provide a detailed exposition of FPGAgent. The experimental setup and results are detailed in Sections 4 and 5. Threats to Validity are discussed in Section 6. Related work is introduced in Section 7. Finally, the paper concludes in Section 8 with an overview of future work.

\vspace{-0.6em}

\section{Motivating Example}

Recent LLM-based research for HLS has made steady progress in assessing whether generated HLS code can be compiled, simulated, and synthesized under standard toolchains~\cite{khan2026bench4hls,abi2025hls}. Emerging agent-based workflows further integrate co-simulation and synthesis feedback to iteratively refine and optimize FPGA kernels~\cite{moraru2026laafd,mashnoor2025timelyhls}. However, from the perspective of FPGA deployment, these “front-end” metrics provide only a partial proxy: even if a design passes simulation, it may still fail to produce a deployable FPGA executable, or it may encounter runtime failures during board-level execution. Using a concrete FPGA kernel case study on the Xilinx U200 platform, this section aims to make this gap explicit and to demonstrate the motivation behind FPGAgent. \par

Figure~\ref{fig:motivating example} illustrates an FPGA kernel that passes simulation-based validation but fails when being compiled into an FPGA executable. As shown, the kernel adopts a simple yet aggressive optimization strategy—enforcing a pipeline initiation interval (II) of 1 and fully unrolling all loops. This results in excessive transactional parallelism within a single kernel cycle, which in turn either lengthens the critical path or induces an overly large number of long-distance, cross-region interconnects. The former leads to timing-closure failures, whereas the latter typically causes routing congestion or unroutable designs (as observed in Figure~\ref{fig:motivating example}).\par

This case study indicates that, for HLS designs, being “simulatable/synthesizable” is a necessary condition for “executability on an FPGA with correct board-level behavior,” but it is not sufficient. If LLMs are to effectively assist FPGA engineers—i.e., to rapidly produce HLS implementations that are both functionally correct and reasonably performant—validation confined to simulation environments is far from adequate. Instead, a more low-level, end-to-end framework for generating and validating FPGA executables is required to bridge the gap between task-level specifications and board-level verification.

\vspace{-0.6em}

\section{Methodology}
This section introduces an automated HLS code generation and validation method composed of multiple context-aware agentic modules informed by expert experience and error-driven analysis. \par

\begin{figure*}
    \centering
    \includegraphics[width=0.9\linewidth]{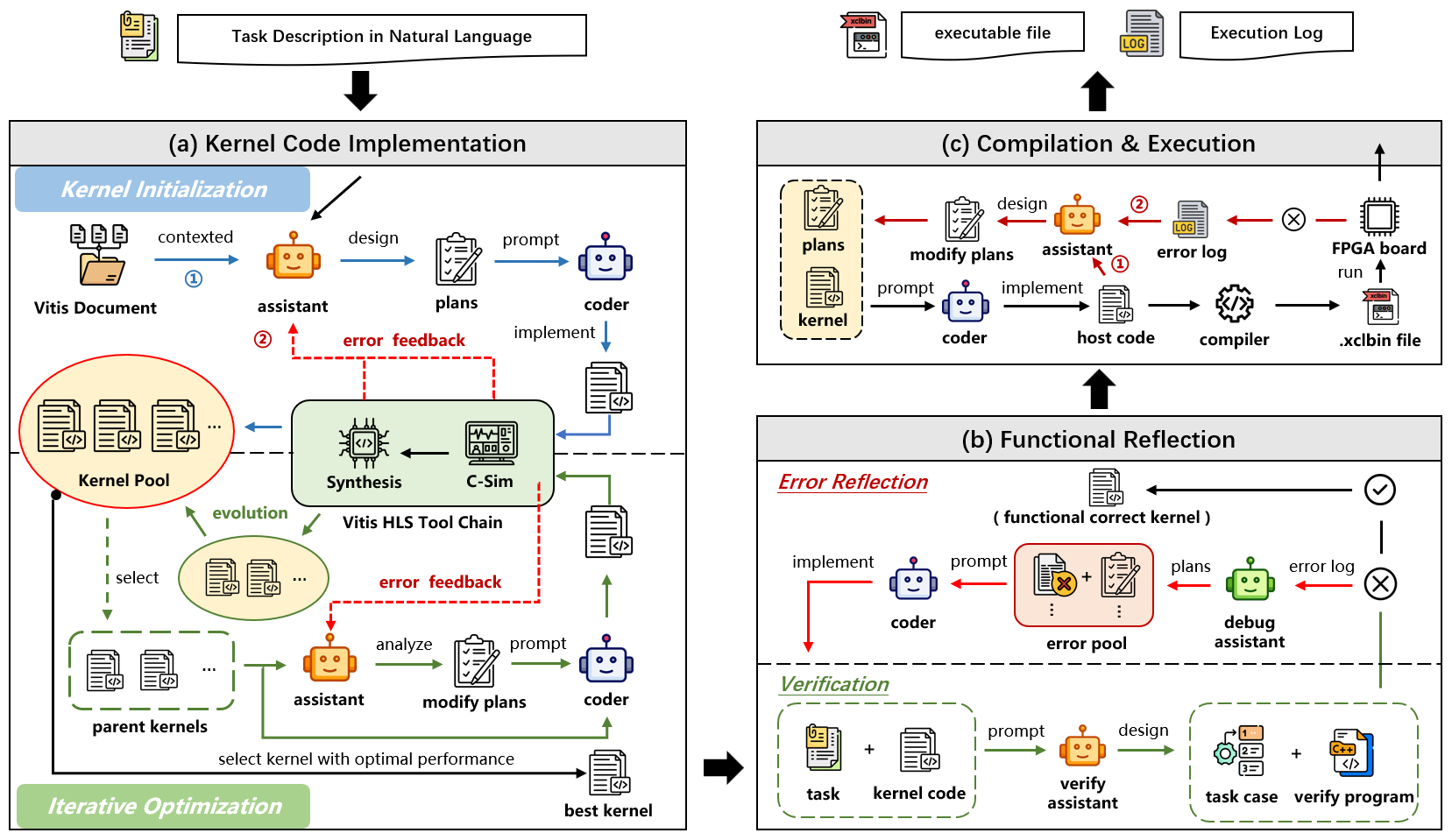}
        \caption{FPGAgent aims to construct an end-to-end framework for HLS code generation and verification, spanning the entire process from design requirements to FPGA-executable programs. (a) Kernel Code Implementation module generates kernel code by employing an optimization framework based on evolutionary algorithms, integrated with a multi-level error feedback mechanism. (b) Functional Reflection module verifies the functional correctness of generated kernel code, analyzes the causes of errors, and iteratively rectifies functional defects encountered. (c) Compilation \& Execution module generates FPGA host code, compiles the code into an FPGA-executable program, and performs verification testing on real hardware environment.}
    \label{fig:Workflow}
\end{figure*}

As illustrated in Figure~\ref{fig:Workflow}, the FPGAgent framework consists of three main stages: (a) generating syntactically correct FPGA kernel code with reliable performance under an evolutionary program search paradigm guided by domain-specific expertise; (b) verifying the functional correctness of the FPGA kernel and repairing it based on verification feedback and analyses of historical failure cases; (c) generating task-oriented FPGA host-side code, followed by joint compilation and execution of the kernel and host code, with validation performed on real FPGA hardware.

\vspace{-0.6em}

\subsection{Kernel Code Implementation}
Kernel Code Implementation is one of the core modules for FPGAgent. Its design is inspired by the heuristic EoH proposed by Liu et al.~\cite{liu2024evolution}, which introduces an evolutionary algorithm–based program search paradigm into HLS design, thereby establishing a solution for generating and performance-optimizing kernel-side HLS code targeting Xilinx FPGAs.
The key idea is to leverage (1) prior knowledge of Vitis HLS design and (2) design experience accumulated by the model during the code-generation process. The framework consists of an assistant and a coder. Algorithm 1 outlines the overall workflow:\par

\begin{algorithm}[htbp]
\caption{Kernel Code Implementation}
\label{alg:kernel-evolution}
\begin{algorithmic}[1]
\Require Task description $Task$, max iterations $I_{\max}$, pool size $P_{\text{size}}$, init factor $co$
\Ensure Best kernel code by performance in the final kernel pool

\State $i \gets 0$ \Comment{iteration counter}
\State $KernelPool \gets \emptyset$
\State $ErrorPool \gets \emptyset$

\While{$i < I_{\max}$}
    \State $TempKernelPool \gets \emptyset$

    \If{$i = 0$}
        \Comment{initial pool generation}
        \State $plans \gets \Call{Assistant.opt\_plan\_selection}{}$

        \While{$|TempKernelPool| < co \cdot P_{\text{size}}$}
            \State $kernel\_code \gets \textsc{Coder.kernel\_generate}($
            \Statex \hspace{\algorithmicindent}\hspace{1.2em} $plans,\ ErrorPool)$

            \State $(pass, logs, perf) \gets \Call{Vitis\_Simulate}{kernel\_code}$

            \If{\textbf{not} $pass$}
                \State $cause \gets \Call{Assistant.analyze}{kernel\_code, logs}$
                \State $ErrorPool \gets ErrorPool \cup \{(kernel\_code, cause)\}$
            \EndIf

            \State $TempKernelPool \gets TempKernelPool \cup \{(kernel\_code, perf)\}$
        \EndWhile

    \Else
        \Comment{evolutionary iteration}
        \While{$|TempKernelPool| < P_{\text{size}}$}
            \State $parents\_kernel \gets \Call{Select}{KernelPool}$
            \State $opt\_plans \gets \Call{Explore/Modifiy}{parents\_kernel, Task}$

            \State $kernel\_code \gets \textsc{Coder.kernel\_generate}($
            \Statex \hspace{\algorithmicindent}\hspace{1.2em} $opt\_plans,\ parents\_kernel)$

            \State $(pass, logs, perf) \gets \Call{Vitis\_Simulate}{kernel\_code}$

            \If{\textbf{not} $pass$}
                \State $cause \gets \Call{Assistant.analyise}{kernel\_code, logs}$
                \State $ErrorPool \gets ErrorPool \cup \{(kernel\_code, cause)\}$
            \EndIf

            \State $TempKernelPool \gets TempKernelPool \cup \{(kernel\_code, perf)\}$
        \EndWhile
    \EndIf

    \State $KernelPool \gets KernelPool \cup TempKernelPool$
    \State $KernelPool \gets \Call{Top\_Psize\_By\_Performance}{KernelPool, P_{\text{size}}}$

    \State $i \gets i + 1$
\EndWhile

\State \Return $\Call{Top\_1\_By\_Performance}{KernelPool}$
\end{algorithmic}
\end{algorithm}

\textbf{Initialization.} Receiving the task description, the initial pool of kernel code is constructed. During module initialization, a set of HLS code optimization strategies extracted from the official AMD Vitis documentation are maintained, where each strategy specifies the optimization rationale, the applicable scope, and several representative examples. Given an HLS design task described in natural language, the assistant retrieves relevant strategies and selects an appropriate sequence of optimizations as contextual information, prompting the coder to generate FPGA kernel HLS code as well as the corresponding kernel design thought. Obtaining the candidate kernel, Vitis HLS tools are invoked to perform hardware simulation, checking syntactic correctness and synthesizability. If the simulation succeeds, the kernel is added to the Kernel Pool. This iterative process continues until the Kernel Pool reaches a size of $co$ × $P_{\text{size}}$ (to enlarge the search space, we set an init factor $co$ to increase the pool size in the first evolutionary round, which was set to 2 in the experiment). Finally, based on the simulation performance (latency and resource usage), the $P_{\text{size}}$ best-performing kernel codes are selected from the kernel pool to form the initial kernel pool.\par

\textbf{Iterative Optimization.} After obtaining the initial kernel pool, iterative evolutionary optimizations are performed over the pool, where each generation conducts parallel searches via multiple heuristic evolution procedures. Specifically, we employ two heuristic operators, namely Exploration and Modification.\par
\textbf{1) Exploration: }First, several kernel codes are selected from the current kernel pool. The assistant then infers new kernel design ideas that are consistent with the task description by analyzing these kernels and their associated design thoughts, and prompts the coder to synthesize new kernel implementations accordingly. \par
\textbf{2) Modification: }In this process, a single kernel code is randomly sampled from the pool. Under the constraint that the kernel’s functionality and interface remain unchanged, the assistant proposes FPGA-oriented optimization strategies, after which the coder applies these strategies to revise the selected kernel code and produce a new candidate kernel.\par
For each newly generated kernel, Vitis HLS tools are invoked to perform hardware simulation and adds only the candidates that pass simulation to the Kernel Pool. Once the Kernel Pool reaches a size of double $P_{\text{size}}$, best-performing kernel codes are selected from the Kernel Pool, which used as the initial pool for the next evolutionary generation. This iterative process continues until the number of generations reaches max generation $I_{\max}$. Finally, kernel code with the best performance is selected from the Kernel Pool to serve as the final FPGA kernel-side HLS code generated by Kernel Code Implementation framework.\par

\begin{figure*}[t]
    \centering
    \includegraphics[width=0.9\linewidth]{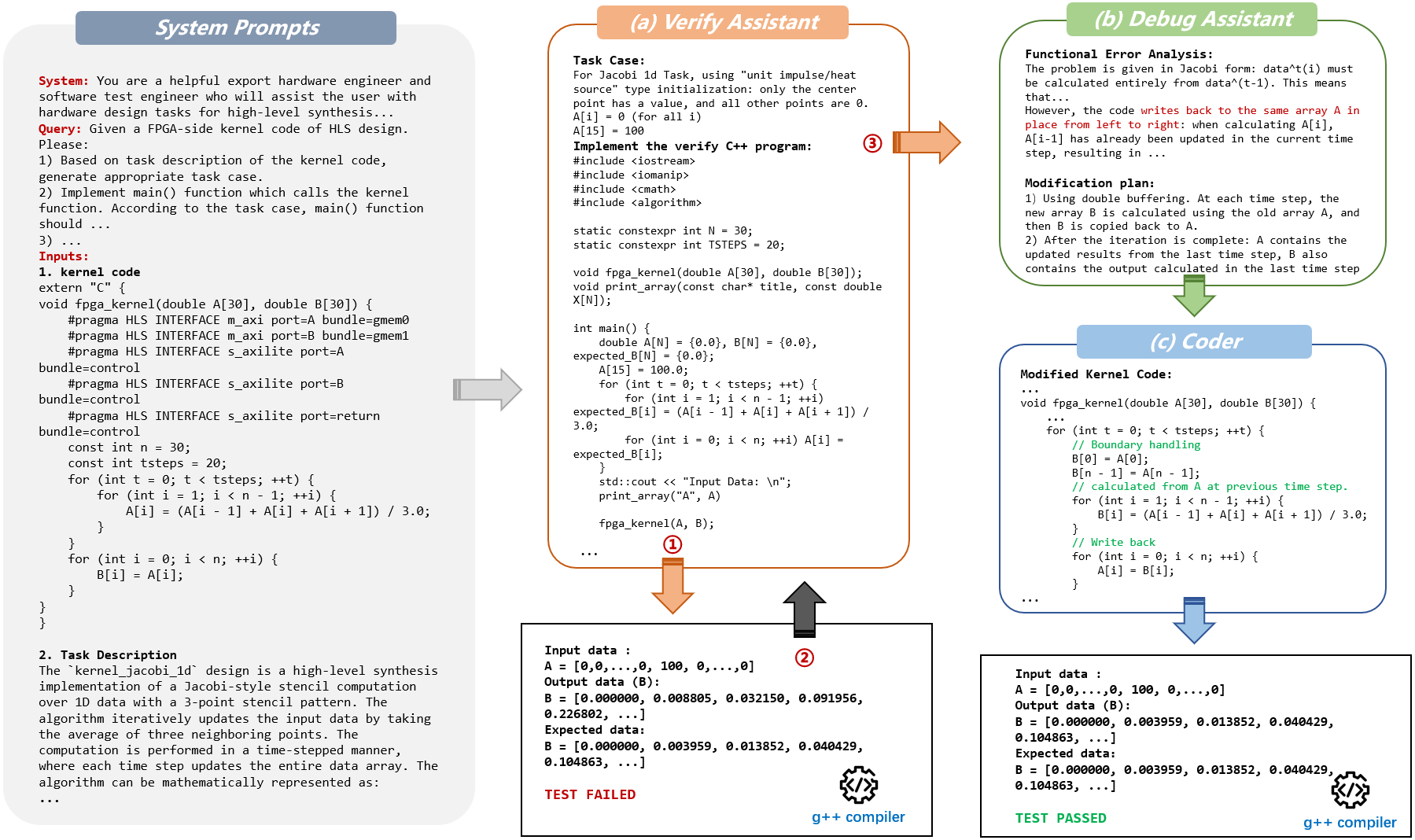}
    \caption{Functional Reflection module verifies the functional correctness of generated kernel code: (a) Verify Assistant generates C++ code and corresponding test cases for functional verification; (b) Debug Assistant analyzes errors in code exhibiting functional faults; (c) Coder performs functional repairs to the code based on the identified root causes of the errors.}
    \label{fig:Functional Reflection}
\end{figure*}

\textbf{Multi-level Error Reflection.} Error Reflection mechanism is designed to ensure the syntactic validity of generated kernel code. After the coder completes a kernel implementation, the system sequentially invokes the Vitis HLS toolchain to perform C-simulation and synthesizability checks. If any step fails, the system collects the corresponding error logs and instructs the coder to revise the kernel accordingly. If the revised code still fails, the assistant analyzes the failure by jointly considering the kernel implementation and the error logs. The system then records the syntactically invalid kernel together with the diagnosed root cause in an Error Pool, thereby maintaining a history of mistakes encountered during HLS generation. In subsequent HLS generation iterations, these error exemplars are explicitly incorporated into the prompt context, helping prevent the coder from repeating the same errors.

\vspace{-0.2cm} 

\subsection{Functional Reflection}
Kernel Code Implementation completes FPGA kernel-side HLS code generation from a high-level task description, producing implementations that are syntactically correct and exhibit reasonable performance. However, the functional correctness of the generated kernels haven't been explicitly validated, which implies that a kernel may compile successfully and execute in an FPGA environment while still producing incorrect outputs. To address this limitation, we propose a Functional Reflection module, which aims to verify functional correctness and repair potential functional defects in the generated kernel code. The Functional Reflection framework comprises a Verifier, a Debugger and a Coder, and its overall workflow is illustrated in Figure~\ref{fig:Functional Reflection}.\par

\textbf{Verification.} Functional Reflection performs C/C++-level functional verification for generated kernel codes. First, the Verify Assistant constructs appropriate input data from the task specification and computes the corresponding expected outputs. Next, the Verify Assistant de-HLSifies the kernel by removing HLS optimization pragmas and directives, yielding C kernel (i.e., the pure C/C++ portion of the kernel). Based on this, the Verify Assistant writes a kernel validation program whose main function is responsible for (1) ingesting the inputs, (2) invoking C kernel, and (3) checking whether the produced outputs match the expected results. Finally, system compiles and executes the validation program using g++ to determine whether the kernel’s output is correct.\par

\textbf{Error Analysis.}
For kernels that fail functional verification, error analysis is performed. The Debug Assistant proceeds by (1) identifying what the major computational stages of the kernel are and what functional modules are required at each stage, as dictated by the task specification; (2) determining where the fault resides and why it occurs through joint analysis of the validation inputs, expected outputs, and observed outputs; and (3) formulating repair strategies that are feasible in practice and recording the functionally incorrect kernel–repair strategy pairs in an Error Pool.\par

\textbf{Iterative Fix.}
After registering these errors, the Error Pool is incorporated into the prompt context to guide the Coder in generating revised kernel implementations by referencing previously observed failure cases and their corresponding fixes. The updated kernel is then submitted to the Verify Assistant for functional validation. This procedure iterates until a kernel passes the functional correctness check.

\begin{figure*}[t]
    \centering
    \includegraphics[width=0.85\linewidth]{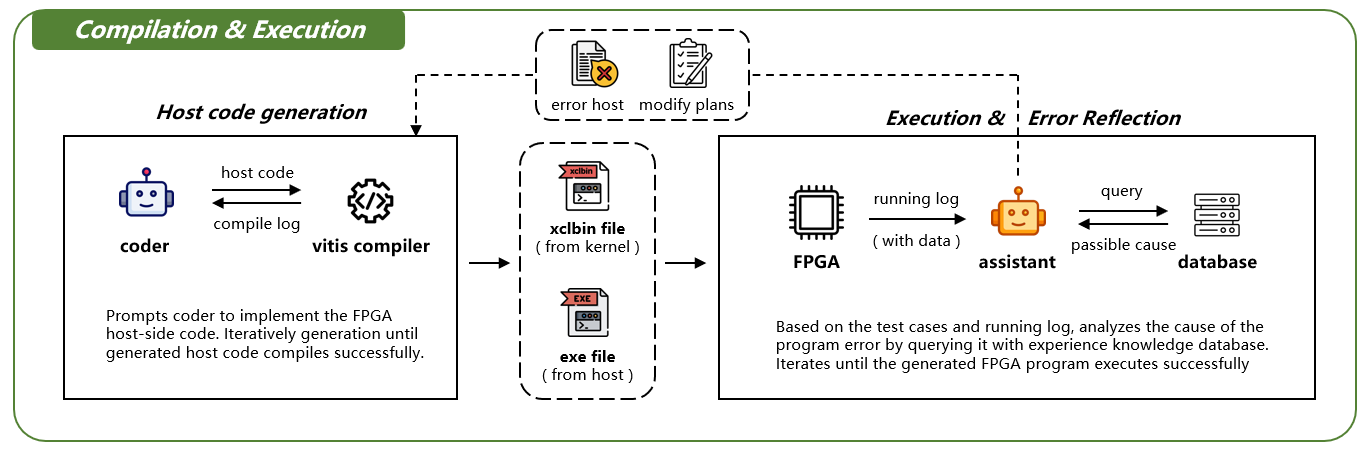}
    \caption{ Compilation \& Execution module completes: (a) FPGA host code generation; (b) compilation of kernel and host code; and (c) execution and verification of the FPGA program.}
    \label{fig:Execution}
\end{figure*}

\vspace{-0.6em}

\subsection{Compilation \& Execution}
Existing LLM-based HLS/HDL code generation approaches predominantly focus on simulation-level validation, e.g., checking synthesizability and functional correctness in a simulated environment. However, substantial discrepancies remain between simulation and real hardware, particularly in terms of data movement, resource management, and runtime orchestration. Consequently, designs that pass simulation may still fail when deployed on physical FPGA platforms. To address this gap, we develop a Compilation \& Execution module. The overall workflow is illustrated in Figure~\ref{fig:Execution}.

After obtaining a functionally correct kernel, the Compilation \& Execution module proceeds as follows:\par

(1) Host code implementation. Prompts the Coder to implement the FPGA host-side code. Specifically, the Coder first derives suitable test cases from the task specification, then instantiates data objects according to the parameter types of the kernel top function, and finally invokes the kernel and checks whether the outputs satisfy the test cases.

(2) Code compilation. Invokes the Vitis toolchain to compile HLS code into an FPGA executable. If compilation fails, the Assistant analyzes the compiler logs, derives actionable fixes, and guides the Coder to regenerate an updated host implementation accordingly.

(3) Program execution. Deploys and executes the resulting program on a physical FPGA platform and verifies that the runtime output meets the requirements.

(4) Error feedback. If runtime execution fails (i.e., the program does not execute correctly on FPGA), the Compilation \& Execution module forwards the error logs to the Assistant. The module maintains an experience-driven database that maps FPGA runtime error logs to root-cause analyses. The Assistant queries this database to diagnose the failure and instructs the Coder to repair the host code. This process iterates until the generated program executes successfully on the FPGA board.

\vspace{-0.6em}

\section{Experiment Setup}
\begin{table*}[t]
\centering
\caption{Comparison of zero-shot (pass@10) and FPGAgent across different base models. FPGAgent delivers consistent improvements across all backbone models, and attaining the highest functional correctness of 92.3\% with Gemini 3.1 Pro.}
\label{tab:main_results}
\begin{tabular}{llccc}
\toprule
Base Model & Method & Synthesizable Rate & Executable Rate & Functional Correctness Rate \\
\midrule
Qwen3-Coder-30B-A3B & zero-shot & 59.0\% (pass@10) & 42.3\% (pass@10) & 30.8\% (pass@10) \\
DeepSeek V3.2 & zero-shot & 74.4\% (pass@10) & 61.5\% (pass@10) & 48.7\% (pass@10) \\
ChatGPT 5.2 & zero-shot & 75.6\% (pass@10) & 64.1\% (pass@10) & 55.1\% (pass@10) \\
Claude Sonnet 4.5 & zero-shot & 84.6\% (pass@10) & 71.7\% (pass@10) & 60.3\% (pass@10) \\
Gemini 3.1 Pro  & zero-shot & 82.1\% (pass@10) & 67.9\% (pass@10) & 56.4\% (pass@10) \\
\midrule
Qwen3-Coder-30B-A3B & FPGAgent & 83.3\% & 75.6\% & 66.7\% \\
DeepSeek V3.2 & FPGAgent & 93.6\% & 91.0\% & 78.2\% \\
ChatGPT 5.2 & FPGAgent & 91.0\% & 87.2\% & 80.8\% \\
Claude Sonnet 4.5 & FPGAgent & 93.6\% & 89.7\% & 84.6\% \\
Gemini 3.1 Pro & FPGAgent & 98.7\% & 97.4\% & 92.3\% \\
\bottomrule
\end{tabular}
\end{table*}

\subsection{Research Questions}
To evaluate the effectiveness of the proposed FPGAgent framework, we conduct a comprehensive study designed to answer the following research questions:

\begin{rqbox}[Research Questions]
\begin{itemize}
\item[\textbf{RQ1: }]\textbf{Can FPGAgent effectively improve the end-to-end success rate of HLS code generation, from task specification to FPGA executable generation and board-level execution?} This question assesses FPGAgent’s capability to produce HLS implementations that run correctly on real FPGA hardware. \textbf{(Main Results)}

\item[\textbf{RQ2: }]\textbf{How much do FPGAgent’s design choices contribute to the overall end-to-end success rate?} This question investigates how the kernel implementation module and the functional reflection module affect the framework’s overall HLS code generation performance.  \textbf{(Ablation Study)}

\item[\textbf{RQ3: }]\textbf{What are the primary sources of end-to-end failures? Which failure can FPGAgent mitigate, and which remain challenging?} This question analyzes how FPGAgent improves HLS generation accuracy and motivates directions for future research. \textbf{(Failure Analysis and Discussion)}
\end{itemize}
\end{rqbox}

\subsection{Benchmark}
Our evaluation is conducted using the HLS-Eval dataset constructed by Stefan et al.~\cite{abi2025hls}. We use it as our primary benchmark because publicly available benchmarks for HLS code generation are still limited, while HLS-Eval offers a relatively large and well-specified evaluation benchmark (further discussion is deferred to the Threats to Validity section). The dataset contains 94 HLS design benchmarks, each accompanied by a task description, a kernel header file, and an interface specification for the kernel code. These benchmarks span a wide range of application domains, including scientific computing/high-performance computing kernels, digital signal processing, cryptography, floating-point computation, and deep learning acceleration. Since the generated HLS designs must ultimately run on Xilinx FPGA hardware in our experiments, we filtered the HLS-Eval benchmarks and made minor adjustments to certain interface definitions, retaining 78 instances that conform to the Vitis HLS design conventions.\par

\vspace{-0.6em}

\subsection{Implementation}
We evaluate five representative general-purpose code generation models, spanning mainstream closed-source and open-source families with different parameter scales, to reflect model choices that are likely to be adopted in practical engineering settings: ChatGPT 5.2, Gemini 3.1 Pro, Claude Sonnet 4.5, DeepSeek V3.2, and Qwen3-Coder-30B-A3B. For inference configuration, given model-specific sensitivity to sampling stochasticity and their recommended settings, we set the temperature to 0.0 for Claude Sonnet 4.5 and DeepSeek V3.2, while using the default temperature for the other models. Except for temperature, all other sampling parameters were kept at platform defaults to better approximate real-world usage and to avoid bias introduced by extensive hyperparameter tuning.\par

All candidate HLS kernels are compiled and synthesized using the Vitis HLS 2022.2 toolchain, which performs HLS C/C++ compilation, checking, and synthesis, and produces intermediate artifacts required for subsequent FPGA deployment. To validate end-to-end executability, we conduct hardware-side experiments on a Xilinx U200 FPGA platform with a target clock frequency of 200 MHz. Under this configuration, the evaluation treats the generated FPGA executable as the final deliverable.

\vspace{-0.6em}

\section{Evaluation}

\subsection{RQ1: Main Results}
The main experiments report the end-to-end HLS code generation performance of FPGAgent across multiple models, and compare it against zero-shot generation. Our primary objective is to determine whether, under varying model capabilities and inference characteristics, FPGAgent can consistently improve end-to-end outcomes in terms of synthesizability, executability, and the final functional correctness.\par

We evaluate FPGAgent with different closed-source and open-source models. To disentangle the backbone model’s inherent generation ability from the systematic gains introduced by the FPGAgent framework, we consider two evaluation modes for each model: (1) Zero-shot, where the task specification is directly provided to the model to generate both the HLS kernel and host code, followed by compilation and execution of the generated design; and (2) FPGAgent, where the same task specification is used but the full generation and verification workflow of FPGAgent is enabled.\par

To cover different stages from HLS code to FPGA executable verification, we set three progressive metrics:(1) Synthesizable Rate: The generated code can be correctly synthesized by the toolchain, reflecting its implementability at the HLS level; (2) Executable Rate: The generated code can be compiled into an FPGA executable file (xclbin) and executed successfully, reflecting its engineering feasibility in the end-to-end build chain; (3) Functional Correctness Rate: Under the premise of execution, the executable file output meets the input task specifications, reflecting the final end-to-end functional satisfaction.\par

\begin{table*}[t]
\centering
\caption{Ablation study of different components in FPGAgent.}
\label{tab:ablation}
\begin{tabular}{lccccc}
\toprule
Variants & \makecell{Kernel Evolution\\Optimization} & \makecell{Functional\\Reflection} & Synthesizable Rate & Executable Rate & \makecell{Functional\\Correctness Rate} \\
\midrule
\textsc{FPGAgent\_base}    & \xmark & \xmark & 59.0\% & 47.4\% & 37.2\%  \\
\textsc{FPGAgent\_eoh}     & \cmark & \xmark & 80.1\% (+17) & 66.7\% (+15) & 48.7\% (+9) \\
\textsc{FPGAgent\_reflect} & \xmark & \cmark & 57.7\% (-1) & 50.0\% (+2) & 46.2\% (+7) \\
\textsc{FPGAgent\_full}    & \cmark & \cmark & \textbf{83.3\% (+19)} & \textbf{75.6\% (+22)} & \textbf{66.7\% (+23)} \\
\bottomrule
\end{tabular}
\end{table*}

In the experiment, the kernel pool size in the FPGAgent kernel code implementation was set to 2 (the initial kernel pool size was 4), and the number of iteration rounds was set to 3. To ensure fairness at the sampling level, we calculated the pass@10 success rate of the above metrics for zero-shot generation. Table~\ref{tab:main_results} shows our experimental results.

Overall, despite being a training-free framework, FPGAgent achieves consistent and robust performance improvements across all backbone models. Compared with the zero-shot baseline, FPGAgent improves synthesizability by 16.9\% on average, executability by 26.7\%, and functional correctness by 30.6\%. Notably, while FPGAgent yields gains for both closed-source and open-source models of different scales, the magnitude of improvement varies by model. With Gemini 3.1 Pro, FPGAgent achieves the highest functional correctness of 92.3\%; for other large-scale models, the functional correctness reaches 78–85\%; and for the medium-sized open-source model Qwen3-Coder-30B-A3B, FPGAgent still attains 66.7\% functional correctness. These results indicate that the framework is not only effective for high-capability proprietary models, but also substantially enhances the practicality and feasibility of medium-sized open-source models for HLS generation.\par

Meanwhile, we observe that zero-shot HLS generation exhibits substantial variability in performance: it may occasionally produce implementations comparable to human-written code, but can also yield designs with excessive latency or resource consumption. In contrast, FPGAgent improves generation stability—over successive iterations, the resulting HLS designs tend to progressively improve and eventually converge to a reasonable performance regime. This observation partially explains why FPGAgent increases both synthesizability and executability.\par

In summary, FPGAgent significantly improves end-to-end HLS generation quality under diverse model configurations: it increases the proportion of synthesizable designs, improves the likelihood of successfully building xclbin artifacts and executing them on hardware, and ultimately yields higher functional correctness. This demonstrates that FPGAgent’s end-to-end workflow effectively advances HLS code generation beyond simulation and synthesis validation toward runnable deployment in real FPGA environments.

\vspace{-0.5em}

\subsection{RQ2: Ablation Study}
In the ablation experiment, we analyze the contribution of each key module of the FPGAgent to the end-to-end HLS code generation effect. We constructed four variants to isolate two key components of FPGAgent: the kernel evolution optimization module (\textsc{Eoh}) and the kernel functional reflection module (\textsc{Reflect}).\par
\textsc{FPGAgent\_base}: Serving as an ablation baseline, removes \textsc{Eoh} and \textsc{Reflect}, retaining only the zero-shot kernel generation with RAG (containing HLS optimization instructions/official documentation snippets in the context), and the subsequent host code generation and FPGA program execution modules; \textsc{FPGAgent\_eoh}: Adds \textsc{Eoh} to \textsc{FPGAgent\_base}. This module employs an optimization strategy based on evolutionary algorithms and utilizes a multi-level feedback mechanism to guide the model in learning design expertise; {FPGAgent\_reflect}: Adds \textsc{Reflect} to \textsc{FPGAgent\_base}. This module verifies the functional correctness of the kernel, locates potential errors and derives repair strategies; \textsc{FPGAgent\_full}: The complete method, enabling both \textsc{Eoh} and \textsc{Reflect}.\par
The ablation experiment used Qwen3-Coder-30B-A3B as the base model. Since \textsc{Eoh} triggers multiple kernel generation, to ensure fair comparisons between different variants, we used pass@10 as the primary evaluation metric for \textsc{FPGAgent\_base} and \textsc{FPGAgent\_reflect} to match the multi-candidate search characteristics introduced by \textsc{Eoh}. \par

(1) Basic Capability of the Agentless Workflow:
\textsc{FPGAgent\_base} successfully solves 29 instances in the test set, outperforming pure zero-shot generation. This result suggests that incorporating design guidelines and pragma-based optimization directives from the official Vitis HLS documentation as contextual information can improve HLS generation success to a certain extent. \par
(2) Improved Capability of the Kernel Evolution Optimization Component:
After introducing \textsc{Eoh}, \textsc{FPGAgent\_eoh} exhibits substantial gains in synthesizability, executability, and functional correctness. Specifically, it solves 9 more instances than \textsc{FPGAgent\_base} (a relative improvement of 31.0\%), indicating that iterative evolutionary refinement is crucial for improving the quality of generated HLS code (as shown in Table~\ref{tab:ablation}). \par
(3) Improved Capability of the Kernel Functional Reflection Component:
With \textsc{Reflect} enabled, \textsc{FPGAgent\_reflect} effectively increases the functional correctness of generated kernels without degrading synthesizability or executability. In particular, it solves 8 more instances than \textsc{FPGAgent\_base} (a relative improvement of 27.6\%) (see Table~\ref{tab:ablation}). This aligns with the module’s design intent: by localizing and repairing functional defects, \textsc{Reflect} helps the model overcome the common failure mode where code is runnable but produces incorrect results. \par
(4) Full Capability of the Complete Workflow:
Although \textsc{Eoh} and \textsc{Reflect} each provide significant improvements, the results indicate that relying on a single component is insufficient to fully realize FPGAgent’s potential. Using \textsc{Eoh} only (\textsc{FPGAgent\_eoh}): without strong feedback on functional correctness during optimization, the system may produce kernels that compile and perform well but are functionally incorrect, which limits end-to-end performance. Using \textsc{Reflect} only (\textsc{FPGAgent\_reflect}): its verification is primarily conducted at the C/C++ level and is not tightly coupled to the semantics of the backend hardware implementation, making it difficult to ensure hardware-level correctness and, consequently, end-to-end executability. In contrast, the complete system \textsc{FPGAgent\_full} solves 52 benchmark instances, substantially outperforming all other variants. \par
Overall, the ablation results indicate that: (1) adding design guidelines and pragma-based optimization directives to the prompt context improves HLS generation success; (2) kernel evolution optimization module (\textsc{Eoh}) is a key driver for synthesizability, executability, and build robustness; (3) kernel functional reflection module (\textsc{Reflect}) primarily improves functional correctness; (4) both modules are indispensable and act synergistically in FPGAgent, yielding the highest end-to-end success rate.\par

\vspace{-0.5em}

\subsection{RQ3: Failure Analysis and Discussion}
\begin{figure}
    \centering
    \includegraphics[width=0.9\linewidth]{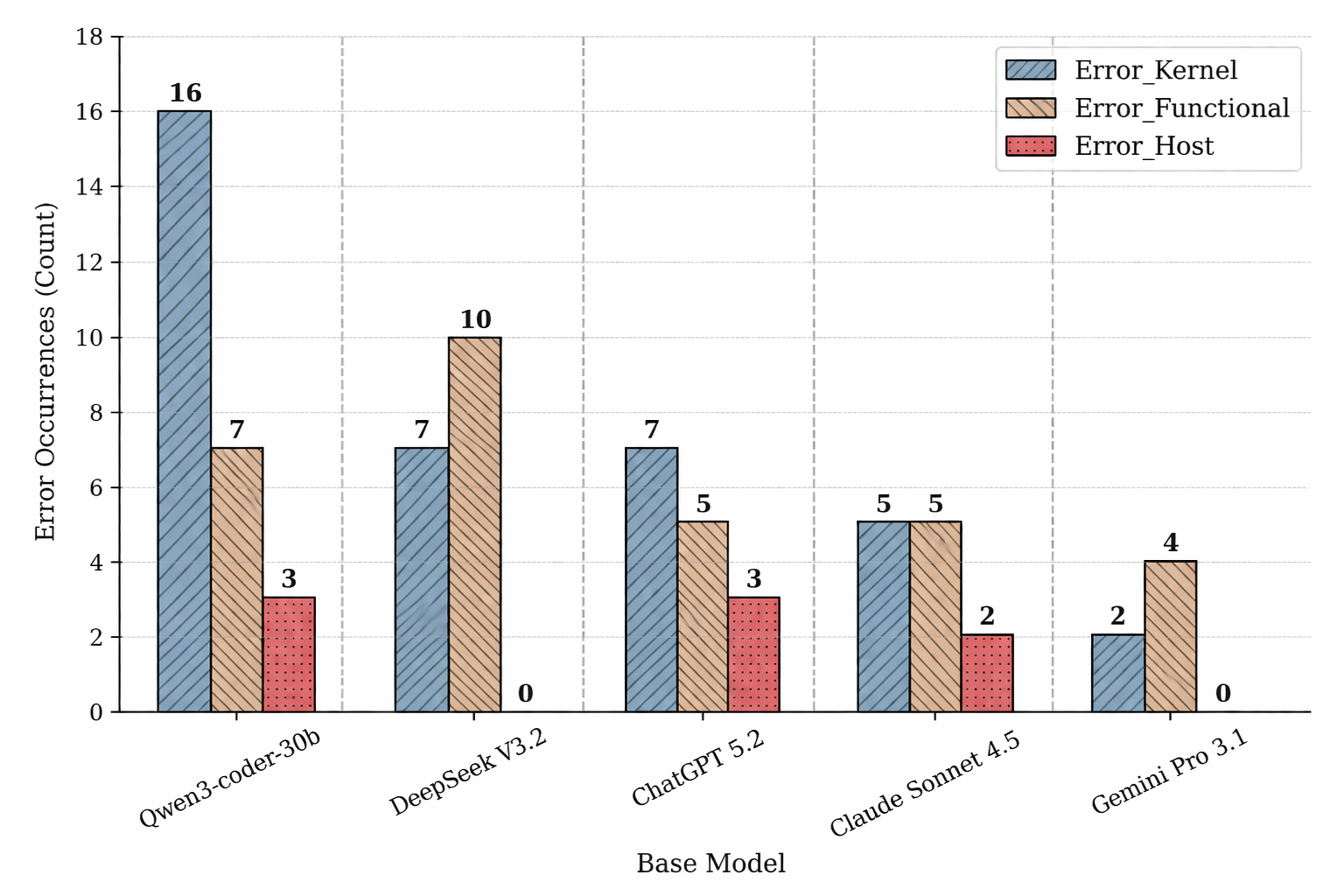}
    \caption{Error Distribution Across Different Models}
    \label{fig:Error Distribution}
\end{figure}

To further investigate the root causes of HLS generation failures and to provide a principled explanation for why FPGAgent improves generation accuracy, we perform a statistical analysis of end-to-end failures in producing FPGA executables (i.e., xclbin + host executable) observed during the experiments. We categorize these failures into three types—Error\_Kernel, Error\_Functional, and Error\_Host—and summarize their distributions across different backbone models in Figure~\ref{fig:Error Distribution}.\par

\textbf{1) Error\_Kernel:} Refers to errors/non-compliance in the generated kernel code, causing the end-to-end process to fail. It manifests in two typical scenarios:
(1) All candidate kernels in the kernel pool fail the simulation test (i.e., Vitis HLS hw\_emu mode), indicating defects at the syntactic or functional level;
(2) The selected best kernel can pass hw\_emu simulation but cannot be successfully compiled into an FPGA executable (xclbin), suggesting unreasonable design choices at the timing, micro-architectural, or implementation level.
For the latter case, the dominant cause is not syntax errors but inefficient kernel implementations, which may trigger timing-closure failures or resource exhaustion during the hardware build stage, ultimately preventing deployment.

\begin{txtbox}[Causes of the "Error\_Kernel"]
Insufficient task understanding and programming capability of the backbone model is a major cause to Error\_Kernel, and the gap across models is pronounced.
\end{txtbox}
Our statistics indicate that when using Qwen3-Coder-30B-A3B as the kernel-generation model, 18 instances fail due to Error\_Kernel; in contrast, for stronger frontier LLMs (DeepSeek V3.2, ChatGPT 5.2, Gemini 3.1 Pro, and Claude Sonnet 4.5), the numbers decrease to 7, 7, 2, and 5, respectively (Figure~\ref{fig:Error Distribution}). This disparity suggests that, for HLS generation, the backbone model’s fundamental capability substantially affects the synthesizability/implementability of generated kernels.\par

Moreover, we observe that the evolutionary kernel-optimization component in FPGAgent markedly reduces the incidence of Error\_Kernel. Under zero-shot generation with Qwen3-Coder-30B-A3B, Error\_Kernel occurs 29 times; when enabling \textsc{FPGAgent\_eoh}, the count drops to 17 (see Table~\ref{tab:error_analysis}). This result indicates that, for open-source models with weaker base capabilities, iterative refinement and optimization strategies can effectively improve the synthesizability/implementability of generated kernels.\par

\textbf{2) Error\_Functional:} Denotes cases where the generated HLS design can be compiled into an xclbin and executed on FPGA hardware, yet the produced outputs deviate from the expected results.\par

\begin{txtbox}[Notable “Hardware Mismatch” Failure]
Some generated HLS kernels produce correct outputs in simulation but produce inconsistent outputs after deployment on real FPGAs. This failure mode motivates our hardware-aware prompting and reinforces the need for end-to-end executable verification.
\end{txtbox}
Our analysis suggests that this discrepancy mainly stems from a semantic gap between software-level code behavior and hardware-level implementation. In particular, stack-allocated arrays and related memory accesses may behave as expected under standard C/C++ semantics, but after HLS inference and optimization into on-chip memories such as BRAM or URAM, the resulting hardware behavior may deviate from simulation. Similarly, some HLS pragmas, such as dataflow, impose structural constraints that are not fully captured by conventional simulation; when used improperly, they can also trigger hardware-only functional failures.\par

To mitigate such “hardware mismatch” failures, FPGAgent explicitly injects a set of empirically collected canonical error cases into the prompt context during kernel generation, including erroneous code snippets, root-cause analyses, and corresponding repair strategies, to discourage the model from repeating mistakes. Our statistics indicate that this strategy is effective: for Qwen3-Coder-30B-A3B, after adding these prompts, inconsistency occurs in only one test case, compared to ten before the prompts were introduced.\par

\begin{txtbox}[Functional Repair Bottleneck]
The Functional Reflection module does not fully repair all failing instances, especially on some complex tasks. Such cases constitute a major bottleneck for further improving end-to-end success rate.\par
\end{txtbox}
We hypothesize that, for more complex benchmarks, the model may be unable to accurately localize the root cause using only input–output discrepancies (e.g., issues involving boundary conditions, state dependencies, or subtle data-layout effects), which limits its ability to derive effective fixes. These motivate an important direction for future work: integrating more interpretable and fine-grained feedback mechanisms to enhance the model’s capability to repair functional defects in complex task settings.

\textbf{3) Error\_Host:} Refers to failures that occur during FPGA program execution due to defects in the host-side code or misconfigurations of the runtime. Such errors are tightly coupled to engineering details, including the Vitis runtime, Xilinx platform configuration, and host–kernel interface binding (e.g., the definition of kernel-argument buffer objects and the choice of data-transfer mechanisms). These issues typically arise during compilation and runtime execution, and their manifestations are highly dependent on the target deployment environment.\par

To address Error\_Host, FPGAgent adopts a combined strategy of contextual constraints and error-feedback-driven repair, including:
(1) explicitly injecting canonical failure cases and correct reference implementations into the prompt context to steer the model away from high-risk coding patterns;
(2) when execution failures occur, diagnosing root causes from runtime logs and feeding actionable repair suggestions back to the model to iteratively fix the host program.
Experimental results demonstrate the effectiveness of this strategy: as shown in Figure~\ref{fig:Error Distribution}, we can find that Error\_Host becomes rare in the final outcomes.

\begin{table}[t]
\centering
\small
\caption{Error analysis of FPGAgent variants.}
\label{tab:error_analysis}
\begin{tabular}{lcccc}
\toprule
Variant & Total & Kernel & Functional & Host \\
\midrule
\textsc{FPGAgent\_base}    & 49 & 29 & 15 & 5 \\
\textsc{FPGAgent\_eoh}     & 40 & 17 & 19 & 4 \\
\textsc{FPGAgent\_reflect} & 42 & 31 & \textbf{5} & 6 \\
\textsc{FPGAgent\_full}    & \textbf{26} & \textbf{16} & 7 & \textbf{3} \\
\bottomrule
\end{tabular}
\end{table}

In summary, FPGAgent reduces kernel non-implementability via its evolutionary optimization module, mitigates board-level semantic mismatches through functional error feedback and contextualized error exemplars, and lowers the platform-dependent integration failures through log-driven host repair. These mechanisms provide a principled explanation for the observed improvements in end-to-end success rate and functional correctness for  HLS code generation.

\section{Threats to Validity}

\textbf{Internal Validity.} Our primary experiments are conducted on the HLS-Eval benchmark, which introduces a potential threat to validity: reliance on a single benchmark may bias the reported results. We focus on HLS-Eval because well-established benchmarks for LLM-assisted HLS remain limited; among the available options, HLS-Eval provides a well-specified interface and non-trivial scale, and it spans a broad and diverse set of common HLS application domains—including scientific computing/HPC kernels, digital signal processing, cryptography, floating-point computation, and deep learning acceleration—which partially mitigates the uncertainty induced by a single task distribution. We also note that more recently studies have begun to introduce larger-scale benchmarks, or benchmarks that emphasize end-to-end evaluation, specifically for the “LLM-assisted HLS” setting (e.g., Bench4HLS~\cite{khan2026bench4hls}). In future work, we plan to leverage these benchmarks to further validate the performance of FPGAgent.\par

\textbf{External Validity.} The end-to-end generation of FPGA executables may introduce tight coupling between the framework and a particular software/hardware stack: We acknowledge that executable generation is inherently hardware-dependent—especially for the host-side code, whose implementation patterns often depend on the target device and the configuration of the runtime stack. Consequently, when porting to different environments, it may be necessary to adapt the prompting strategy or interface templates used for host-code generation. Meanwhile, we emphasize that a central objective of FPGAgent is to improve the quality of kernel-side HLS generation. The kernel’s functional logic is directly determined by the task specification; accordingly, our kernel-generation procedure strictly adheres to Vitis HLS design principles and synthesizability constraints. As a result, platform dependence is primarily concentrated in the host-code generation/compilation stage, whereas the kernel generation and verification mechanism is broadly applicable across platforms.\par

\textbf{Efficiency and Cost Validity.} FPGAgent employs an iterative architecture that combines evolutionary search with multi-level error feedback. Compared to the zero-shot or few-shot generation paradigm commonly used with LLMs, this design may incur higher time overhead: In our end-to-end workflow, we observe that the dominant bottleneck is neither LLM inference nor kernel simulation/validation, but rather the hardware build stage for generating the xclbin bitstream. This stage typically takes approximately 3–5 hours, which is substantially longer than the tens of minutes required for kernel synthesizability checks and other lightweight steps. To control the overall time cost, we adopt a one-shot xclbin generation strategy for each test case: before entering the most expensive xclbin build stage, we perform low-cost synthesis and screening at the kernel level to maximize candidate quality and success likelihood, thereby avoiding the prohibitive overhead of repeatedly rebuilding hardware across multiple trials.

\section{Related Work}
\textbf{Benchmark Building. }A major line of recent work focuses on benchmarking and evaluation infrastructure for LLM-generated HLS code. HLS-Eval and Bench4HLS~\cite{abi2025hls,khan2026bench4hls} establish end-to-end evaluation frameworks that cover parsing, compilation, simulation, and synthesis. These benchmarks are important because they move evaluation beyond text similarity and toward executable correctness and synthesizability. However, most evaluations still stop at the toolchain level (e.g., C-simulation/co-simulation/synthesis reports), rather than validating the generated designs in a real FPGA deployment workflow~\cite{liu2024chatchisel,khan2026bench4hls}.\par

\textbf{Indirect Generation. }Another active direction is LLM-driven HLS generation, refactoring, and repair. Prior work has explored converting software-oriented C/C++ into HLS-compatible implementations, refactoring code for HLS friendliness, and repairing non-synthesizable programs via LLM-guided transformations~\cite{xiong2024hlspilot,xu2025hlsrewriter,xu2024automated}. These methods demonstrate that LLMs can automate a substantial fraction of the manual engineering effort required for HLS adoption. Related efforts also show that HLS can serve as a bridge for downstream RTL generation, where LLMs first generate HLS-friendly C/C++ and then rely on Vitis HLS to produce RTL, improving pass rates over direct RTL generation~\cite{liao2024llms,vijayaraghavan2024chain,huang2024towards}. A key limitation of these approaches is that they do not support end-to-end code generation: the design specification must first be translated into C/C++ code, which can further introduce additional design and verification overhead.\par

\textbf{Performance Optimization. }A complementary research thread targets optimization and directive/pragma generation, which is crucial for HLS quality-of-results (QoR). Recent work uses LLMs (sometimes combined with graph models, retrieval augmentation, or iterative search) to insert or optimize HLS directives such as pipelining, unrolling, and array partitioning~\cite{prakriya2025lift,yao2025high,xu2024optimizing}. These approaches significantly improve latency/resource tradeoffs, but they are often evaluated primarily on synthesis-level QoR metrics. In contrast, fewer works examine whether QoR-driven transformations remain robust under a stricter correctness criterion that includes executable hardware behavior on a target FPGA platform.\par

\textbf{Agentic Workflow. }More recent systems have incorporated agentic workflows into the high-level synthesis (HLS) pipeline. In such workflows, large language models (LLMs) iteratively refine and optimize code by leveraging feedback signals from compilers, simulators, and synthesis logs~\cite{moraru2026laafd,mashnoor2025timelyhls}. These approaches align well with established HLS toolchains (e.g., Vitis HLS) and support multi-round repair and optimization, demonstrating promising potential. Nevertheless, existing work exhibits notable limitations. Timely HLS~\cite{mashnoor2025timelyhls} has been evaluated across multiple FPGA architectures, yet its validation remains confined to the Vivado simulation level, with comparatively limited consideration of board-level execution, deployment readiness, and end-to-end runnability. LAAFD~\cite{moraru2026laafd} validates HLS performance on the xcu250-figd2104-2L FPGA platform; however, it primarily focuses on translating and optimizing HLS from pre-existing C++ implementations, and it lacks experimental validation on well-established benchmark.

\section{Conclusion and Future Work}
In this paper, we present FPGAgent, an end-to-end framework for generating FPGA executables through the coordinated collaboration of multiple AI agents. Given a high-level task specification, FPGAgent can automatically produce HLS code and invoke the industry-standard HLS toolchain to complete simulation, synthesis, and final executable generation. The framework incorporates an evolutionary-search-driven module for FPGA kernel generation, and further introduces an error analysis and feedback module to verify and improve the functional correctness of the generated code. Experimental results demonstrate that FPGAgent substantially increases the accuracy of HLS code generation: compared with zero-shot baselines, it achieves an average improvement of 30.6\%, and attains up to 92.3\% accuracy when using Gemini 3.1 Pro.\par

Future work will primarily focus on improving kernel performance—specifically, reducing latency and improving resource utilization—while preserving executability and functional correctness. In addition, we plan to explore task-specific fine-tuning for HLS, with the goal of enabling more accurate, efficient, and high-performance HLS code generation.

\section{Data Availability Statement}
The source code, scripts, benchmark instance list, and documentation needed to reproduce the main results will be available upon the acceptance of this paper. Due to hardware/toolchain dependencies (Xilinx FPGA board and Vitis HLS environment), full end-to-end hardware execution may require specialized resources.


\bibliographystyle{ACM-Reference-Format}
\bibliography{references}

\end{document}